\documentclass[conference]{IEEEtran}
\usepackage{graphicx}
\usepackage{adjustbox}
\usepackage{cite}
\usepackage{makecell}
\usepackage{array}
\IEEEoverridecommandlockouts
\usepackage{ctable}

\begin{document}

\title{Cyber Threat Intelligence Model: An Evaluation of Taxonomies, Sharing Standards, and Ontologies within Cyber Threat Intelligence}

\author{
\IEEEauthorblockN{Vasileios Mavroeidis}
\IEEEauthorblockA{University of Oslo\\ Norway\\ vasileim@ifi.uio.no}
\and
\IEEEauthorblockN{Siri Bromander}
\IEEEauthorblockA{mnemonic\\ University of Oslo\\ Norway \\ siri@mnemonic.no}
}

\maketitle

\begin{abstract}
Cyber threat intelligence is the provision of evidence-based knowledge about existing or emerging threats. Benefits from threat intelligence include increased situational awareness, efficiency in security operations, and improved prevention, detection, and response capabilities. To process, correlate, and analyze vast amounts of threat information and data and derive intelligence that can be shared and consumed in meaningful times, it is required to utilize structured, machine-readable formats that incorporate the industry-required expressivity while at the same time being unambiguous. To a large extent, this is achieved with technologies like ontologies, schemas, and taxonomies. This research evaluates the coverage and high-level conceptual expressivity of cyber-threat-intelligence-relevant ontologies, sharing standards, and taxonomies pertaining to the who, what, why, where, when, and how elements of threats and attacks in addition to courses of action and technical indicators. The results confirm that little emphasis has been given to developing a comprehensive cyber threat intelligence ontology, with existing efforts being not thoroughly designed, non-interoperable, ambiguous, and lacking proper semantics and axioms for reasoning.\footnote{This paper is an updated version of DOI: 10.1109/EISIC.2017.20 and includes language enhancements. The changes in no case have affected the paper's scope, analysis, and derived conclusions. The original version of this research is included in the proceedings of the 2017 European Intelligence and Security Informatics Conference (EISIC).\\Revision: Vasileios Mavroeidis\\Date: August 2023}\\
\end{abstract}
	

\section{Introduction}
\label{sec:intro}

Defenders utilize different tools to prevent, detect, and disrupt adversarial operations. However, the increasing capability, persistence, and complexity of the attacks have made traditional defense approaches insufficient.

Organized cybercrime is at its peak, with PwC's 2016 global economic crime survey \cite{pwcArticle} reporting that, on average, organizations have suffered losses of over \$5 million, and of these, nearly a third reported losses of over \$100 million. Juniper Research \cite{juniperResearchArticle} reported that cybercrime will increase the cost of data breaches to \$2.1 trillion globally by 2019, four times the estimated cost of breaches in 2015.
    
Defenders should identify and understand the threats their organizations may face to enhance their security posture. In that respect, threat information sharing could allow one organization's detection to become another's prevention. This practice has achieved a particular maturity, with organizations focusing on generating and sharing highly contextual, accurate, and relevant information, otherwise known as cyber threat intelligence. Threat intelligence is the task of gathering evidence-based knowledge, including context, mechanisms, indicators, implications, and actionable advice about an existing or emerging menace or hazard to assets that can be used to inform decisions regarding the subject's response to that menace or hazard\footnote{https://www.gartner.com/doc/2487216/definition-threat-intelligence}. Organizations rely on cyber threat intelligence to stay threat-informed, identify and understand impending attacks, speed up security operations, and drive and prioritize the implementation of security controls. 

To a large extent, intelligence generation, consumption, and utilization should be supported by automation, which requires leveraging machine-readable representation formats for processing, correlation, and analysis at scale. A type of knowledge representation is ontologies. Ontologies represent knowledge about a particular domain of discourse in a structured manner, leverage logics to perform reasoning, and are flexible and modular so that they can be easily extended, refined, and integrated with other ontologies.

Working towards an ontology for cyber threat intelligence has its challenges. Our research reports the following as the largest barriers to overcome:

\begin{itemize}
    \item Currently, there is little focus on designing ontologies for cyber threat intelligence, particularly ontologies that can support different types of intelligence, namely, strategic, operational, and tactical.
    \item Ambiguity in defined concepts that prevents ontology integration and adoption.
    \item Extensive use of prose and limited utilization of existing taxonomies, vocabularies, and common frameworks and languages undermine the querability of the knowledge bases and graphs and result in a lack of interoperability and reasoning.
    \item Lack of relationships between concepts that could otherwise allow deriving comprehensive explainable views.
    \item Minor use of semantic axioms and constraints in support of consistency checking and information inference. 
\end{itemize}

This article evaluates taxonomies, sharing standards, and ontologies relevant to the task of creating a comprehensive cyber threat intelligence ontology. To achieve that, we created the cyber threat intelligence model to distinguish different types of information in favor of representing the five W's and one H (who, what, why, where, when, how) of threats and threat operations in addition to technical indicators and courses of action. We use the model to determine the expressivity and coverage of the identified taxonomies, sharing standards, and ontologies. Finally, we examine shortcomings in current ontological approaches and discuss directions to enhance their quality.

\section{Methodology}
This section introduces two models related to measuring an organization's threat detection maturity and representing cyber threat intelligence. The two models overlay but can satisfy different requirements, as explained in the following two subsections. The cyber threat intelligence model is used to conduct the review presented in this research.

\subsection{The Detection Maturity Level Model - DML}
   
\begin{figure}[!hbt]
\begin{center}
	\includegraphics[width=\columnwidth]{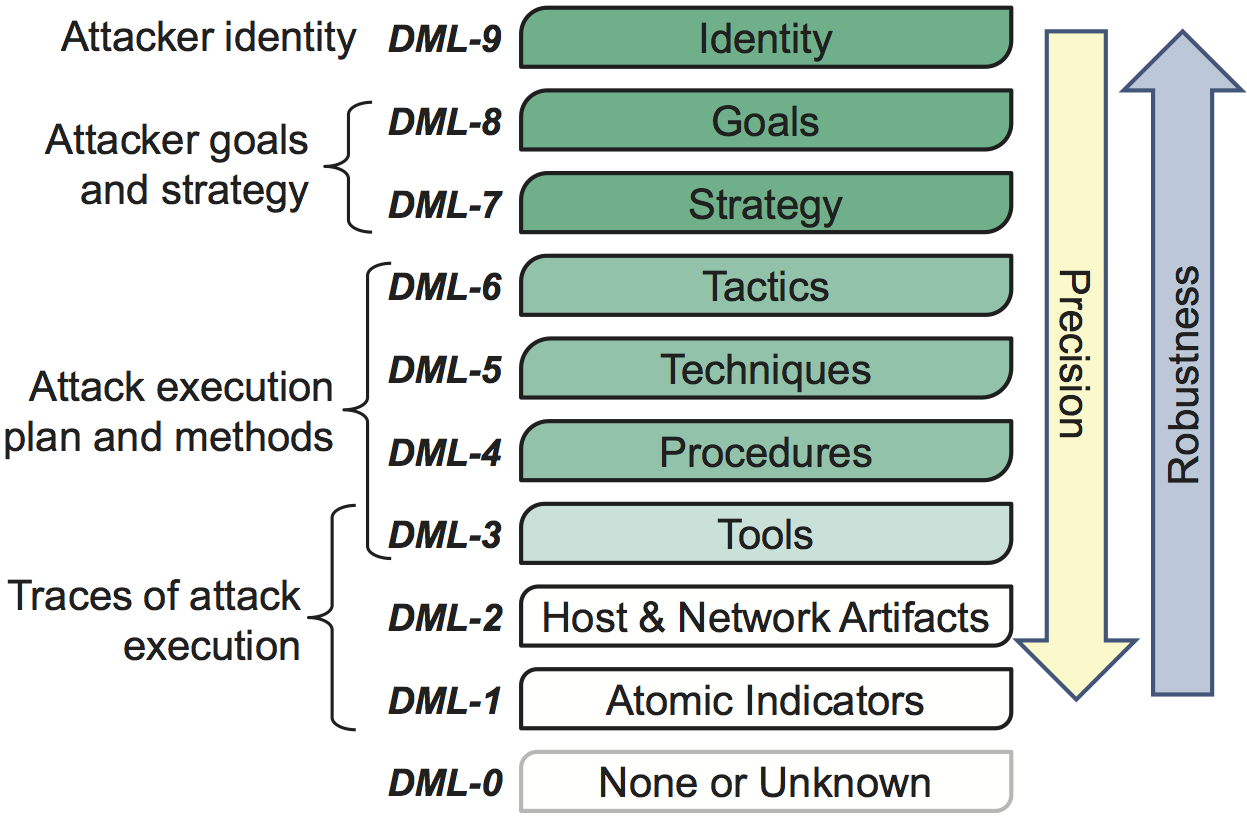}
	\caption{Modified Detection Maturity Level Model\cite{bromandersemantic} \cite{Stillions2014-DML}}
	\label{fig:DML-model}
\end{center}
\end{figure}

Ryan Stillions proposed the Detection Maturity Level (DML) model in 2014 \cite{Stillions2014-DML}. DML is used to describe an organization's maturity regarding its ability to consume and act upon given cyber threat intelligence (Figure \ref{fig:DML-model}).

Detection at the higher levels of DML indicates that an organization is highly intelligence-driven and has established specialized methods for detecting, understanding, and responding to cyber threats more effectively and efficiently. In 2016, we extended this model by adding an additional level (9) "Identity", and presented it for use in the semantic representation of cyber threats \cite{bromandersemantic}.

\subsection{The Cyber Threat Intelligence Model}
The cyber threat intelligence model builds upon and extends \cite{Stillions2014-DML} and \cite{bromandersemantic} to provide a more comprehensive but still high-level view of the different types of information an organization needs access to increase its threat situational awareness. In this research, we utilize this model as a measurement standard to identify the level of expressivity of existing taxonomies, sharing standards, and ontologies.

\begin{figure}[!hbt]
\begin{center}
	\includegraphics[width=\columnwidth]{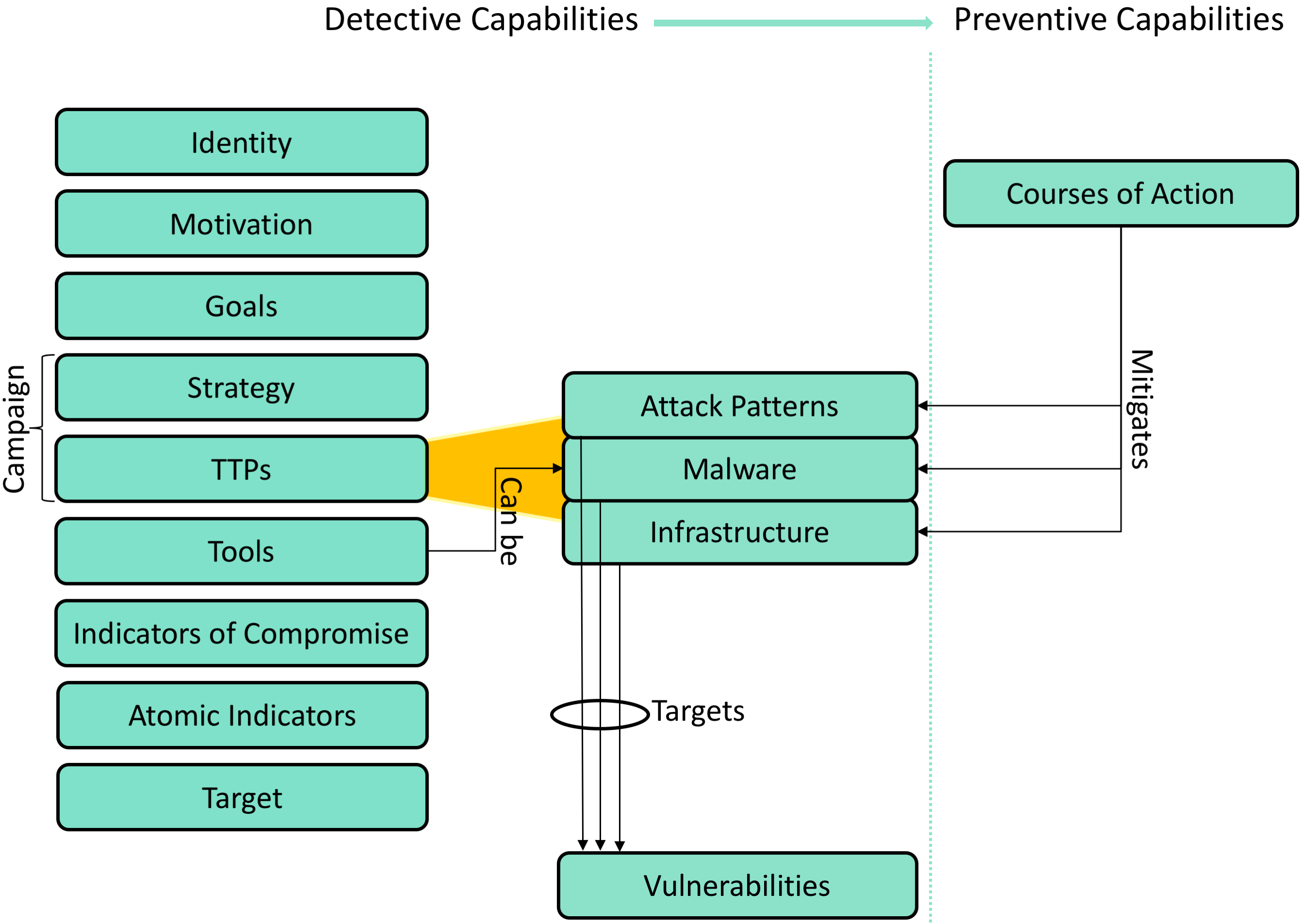}
	\caption{Cyber Threat Intelligence Model}
	\label{fig:CTI-model}
 \end{center}
  \vspace{-5pt}
\end{figure}

The remaining section is devoted to describing the elements comprising the cyber threat intelligence model.

\textbf{Identity:} the identity of a threat actor can be the real name of a person, an organization, a group's affiliates, or a nation-state-backed entity and country. In cases where attribution is not feasible, tracking operations via persona-based profiles (threat actor profiles) also has its benefits as it allows identification of the actor's behavioral characteristics concerning its motivations, goals, capability, and TTPs utilized.

\textbf{Motivation:} can be described as the driving force that enables actions to pursue specific goals. The goals of an attacker may change, but the motivation most of the time remains the same. Knowing a threat actor's motivation narrows down which targets that actor may focus on, helps defenders focus their limited defensive resources on the most likely attack scenarios, as well as shapes the intensity and the persistence of an attack \cite{casey2015understanding}. Examples of motivation can be ideological, geopolitical, and financial.

\textbf{Goals:} according to Fishbach and Ferguson \cite{fishbach2007goal}, "a goal is a cognitive representation of a desired endpoint that impacts evaluations, emotions, and behaviors". A goal consists of an overall end state and the behavior objects and plans needed for attaining it. The activation of a goal guides behaviors. Depending on how the attack is organized, the ultimate goal might not be known to the team executing the attack. The team might only receive direction and a strategy to follow. In current cyber threat intelligence approaches and available knowledge bases, goals are mostly described in prose. However, a goal might be defined as a tuple of two (Action, Object), but work needs to be done to create a consistent taxonomy at a sufficient level of detail \cite{bromandersemantic}. Typical examples of goals are "steal intellectual property", "damage infrastructure", and "embarrass a competitor".

\textbf{Strategy:} is a non-technical high-level description of the planned attack. There are typically multiple ways an attacker can achieve its goals, and the strategy defines which approach the threat agent should follow. 

\textbf{TTPs:} tactics, techniques, and procedures are aimed to be consumed by a more technical audience. TTPs characterize adversary behavior in terms of what they want to achieve technically and how they are doing it.

\textbf{Attack Pattern:} relates to TTPs and describes behavior attackers use to carry out their attacks. 

\textbf{Malware:} relates to TTPs and the capability of the adversary and refers to software inserted into a system with the intent of compromising the target in terms of confidentiality, integrity, or availability.

\textbf{Infrastructure:} describes any system, software service, and any associated physical or virtual resource intended to support an adversarial operation, such as using purchased domains to support command and control (C2), malware delivery 
 and phishing sites.

\textbf{Tools:} attackers install and use tools within the victim's network/infrastructure. Tools encompass both dedicated software developed for malicious reasons and software intended for benign uses (e.g., vulnerability and network scanning, remote process execution, PowerShell) but utilized for malicious purposes, mainly to avoid detection (defense evasion).

\textbf{Indicators of Compromise:} are actionable technical elements/artifacts consumed by cyber security tools to detect intrusions. A good IOC assists threat situational awareness and encompasses contextual information in addition to behavioral (relates to TTPs), computed, or atomic indicators.

\textbf{Atomic Indicators:} the value of atomic indicators is questionable due to their limited context and possible short shelf life. Indicators can include email addresses, domain names, and IPs. Atomic indicators may or may not exclusively represent activity by an adversary.

\textbf{Target:} represents the entity an attack is directed to and can be an organization, a sector, a nation, or an individual.

\textbf{Course of Action:} refers to measures that can be taken to prevent or respond to attacks.

\subsection{Evaluation Criteria}
    
The next sections present and analyze relevant taxonomies, sharing standards, and ontologies found in academic publications/research, open documentation, specifications/standards, and open code/source files. To conduct our study we used the following criteria.
    
\begin{itemize}
\item Use of the cyber threat intelligence model (Figure \ref{fig:CTI-model}) to assess what information/concepts are covered by each work. Table 1 presents the results.
\item Identify integrations (connections) between ontologies, taxonomies, and cyber threat intelligence sharing schemas in support of coverage, expressivity, and interoperability (Sections III \& IV).
\item Identify the level of comprehensiveness and knowledge engineering quality in each ontological effort, including the use of semantic relationships and axioms, and determine whether they can perform deductive reasoning/information inference (Sections IV \& V). 

\end{itemize} 

A number of articles presented contain ontologies that are not described in great detail and offer no reference to the actual ontology files (RDF/OWL), making their evaluation challenging. Furthermore, some open-source ontologies do not provide supporting documentation or a publication.

\section{Taxonomies and Sharing Standards}
This section provides an overview of taxonomies and sharing standards that are used or potentially could be used for cyber threat intelligence representation purposes. We further categorize them into enumerations, scoring systems, and sharing standards.  
    
\subsection{Enumerations}

TAL (Threat Agent Library) \cite{casey2007threat} is a set of standardized definitions and descriptions to represent significant threat agents. The library does not represent individual threat actors; thus, it is not intended to identify people or investigate actual security events. The goal of TAL is to support risk management and specifically identify threat agents relevant to specific assets. In that way, security professionals can proactively build defenses for specific threats.

Tim Casey, in 2015, introduced a new taxonomy for cyberthreat motivations\cite{casey2015understanding}. The taxonomy identifies drivers that cause threat actors to commit illegal acts. Knowing these drivers could indicate the nature of the expected harmful actions.

CVE (Common Vulnerabilities and Exposures) \cite{CVE} is a list of records for publicly known information-security vulnerabilities in software packages.

NVD (National Vulnerability Database) \cite{NVD} is a repository of standards-based vulnerability management data represented using the Security Content Automation Protocol (SCAP). NVD performs analysis on CVEs that have been published in the CVE library. This analysis results in deriving severity metrics (Common Vulnerability Scoring System - CVSS), association with vulnerability types (Common Weakness Enumeration - CWE), and applicability statements (Common Platform Enumeration - CPE), as well as other relevant metadata. 

CPE (Common Platform Enumeration) \cite{CPE} is both a specification and a list. The specification defines standardized machine-readable methods for assigning and encoding names to IT product classes (software and hardware). The CPE enumeration provides an agreed-upon list of official CPE names. 

CWE (Common Weakness Enumeration) \cite {CWE} is a library of software and hardware security weaknesses in support of understanding common flaws and their mitigation.

CAPEC (Common Attack Patterns Enumerations and Characteristics) \cite{CAPEC} is focused on application security and describes the common attributes and techniques employed by adversaries to exploit known weaknesses.

ATT\&CK (Adversarial Tactics, Techniques, and Common Knowledge) \cite{ATTCK} is focused on network defense and describes the operational phases in an adversary's lifecycle, pre- and post-exploit, and details the TTPs adversaries use to achieve their objectives while targeting, compromising, and operating inside a network. It is a valuable resource for understanding adversary behavior and can be used for adversary emulation, behavioral analytics, cyber threat intelligence enrichment, defense gap assessment, red teaming, and SOC maturity assessment. ATT\&CK matrices exist about adversary behavior targeting enterprise environments, mobile, and industrial control systems. Moreover, information on the software adversaries' use, mitigation techniques, procedure examples, and detection recommendations are also available.

\subsection{Scoring Systems}

CVSS (Common Vulnerability Scoring System) \cite{CVSS} is a measurement standard aiming to score vulnerabilities based on their severity. Combined with timely CTI on threats that actively exploit specific vulnerabilities relevant to the organization, CVSS could help identify which vulnerability remediation activities should be prioritized.

CWSS (Common Weakness Scoring System) \cite{CWSS} is part of the CWE project, and it provides a mechanism for scoring and prioritizing software weaknesses using 18 different factors. It is worth mentioning that Mitre's Common Weakness Risk Analysis Framework (CWRAF) can be used in conjunction with CWSS to identify the most important CWEs applying to a particular business and their deployed technologies. The difference between CVSS and CWSS is that the first one scores specific software vulnerabilities (a vulnerability has already been discovered and verified), whereas the latter focuses on scoring software weaknesses and also accounts for incomplete information.

\subsection{Sharing Standards}

A study for existing threat intelligence sharing initiatives concluded that Structured Threat Information eXpression (STIX) is currently the most used standard for sharing threat information \cite{sauerwein2017threat}. STIX is an expressive, flexible, and extensible representation language used to communicate an overall piece of threat information \cite{Barnum2012-STIX}. The STIX architecture comprises different cyber threat information structures such as observables, indicators, incidents, adversary tactics, techniques and procedures, campaigns, intrusion sets, threat actors, targets, and courses of action. Furthermore, STIX was recently redesigned and, as a result, omits some of the objects and properties defined in the first version. The objects chosen for inclusion in the second version represent a minimally viable product that fulfills basic consumer and producer requirements for cyber threat intelligence sharing. Both standards can be used and adapted based on an organization's needs. 
It is worth pointing out that MITRE additionally offers MAEC (Malware Attribute Enumeration and Characterization) \cite{MAEC}, a very expressive malware-sharing language for encoding and communicating high-fidelity information about malware based upon attributes such as behaviors, artifacts, and attack patterns. MAEC can be integrated into STIX or used standalone.

OpenIOC, developed by Mandiant, is an extensible XML schema that enables you to describe technical characteristics that identify a known threat, an attacker's methodology, or other evidence of compromise. The types of information covered directly by OpenIOC are derived mainly from enriched low-level atomic indicators, comprising indicators of compromise, thus covering the IOC category of the cyber threat intelligence model.

\section{Ontologies}
    
Since the work of Blanco et al. \cite{blanco2008systematic} in 2008, we have not identified any overview paper for existing ontologies within the cyber security domain. The authors remark that the scientific community has not accomplished a general security ontology because most of the works are focused on specific domains or the semantic web. The same conclusion was drawn by Fenz and Ekelhart \cite{fenz2009formalizing}. Additionally, Blanco et al. \cite{blanco2008systematic} emphasize the difficulty of combining their identified ontologies due to the non-common interpretation and different terms applied to similar concepts in different ontologies. Our study confirms the same almost 10 years after the study of Blanco et al. \cite{blanco2008systematic}.	
  
While several ontologies relevant to the broader cyber security domain exist, only a few were identified regarding threat information and intelligence representation. For many of them, identifying the mappings to the abstraction layers of the cyber threat intelligence model was challenging due to the fact that they are described only at a very high level and lack code (RDF/OWL) files for further research. The ontologies analyzed hereafter are listed chronologically based on their publication date.

Stefan Fenz and Andreas Ekelhat \cite{fenz2009formalizing} described an information security ontology that can be used to support a broad range of information security risk management methodologies. The high-level concepts of their ontology are derived from the security relationship model described in the National Institute of Standards and Technology Special Publication (NIST SP) 800-12. Concepts to represent the information security domain include threat, vulnerability, control, attribute, and rating. In addition, concepts such as asset, organization, and person are necessary to formally describe organizations and their assets. Lastly, the concept of location is integrated and combined with a probability rating to interrelate location and threat information in order to assign priori threat probabilities. Like other works, the authors have difficulties connecting ambiguous concepts deriving from different standards (e.g., not clear distinction between a threat and a vulnerability).

Wang and Guo \cite{wang2009ovm} proposed an ontology for vulnerability management and analysis (OVM) populated with all existing vulnerabilities in NVD. The basis of the ontology is built on the results of CVE and its related standards, such as CWE, CPE, CVSS, and CAPEC. OVM captures the relationships between the following concepts, which constitute the top level of the ontology: vulnerability, introduction phase (software development life cycle - phases during which the vulnerability can be introduced), active location (location of the software where the flaw manifests), IT product, IT vendor, product category (such as web browsers, application servers, etc.), attack (integration of CAPEC), attack intent, attack method, attacker (human being or software agent), consequence, and countermeasure.

Obrst et al. \cite{obrst2012developing} suggested a methodology for creating an ontology based on already well-defined ontologies that can be used as modular sub-ontologies. In addition, they discussed the usefulness of existing schemas, dictionaries, glossaries, and standards as a form of knowledge acquisition of a domain by identifying and analyzing entities, relationships, properties, attributes, and range of values that can be used in defining an ontology. Their proposed ontology is based on the diamond model of intrusion analysis \cite{caltagirone2013diamond}, which represents the relationships between an adversary (actor), the capabilities of the adversary, the infrastructure or resources the adversary utilizes, and the target of the adversary (victim). The authors state that they first developed the concepts of infrastructure and capabilities, but they are still not in the level of detail they desire. In addition, their current ontology is focused on malware and some preliminary aspects of the diamond model.

A good argumentation for transitioning from taxonomies to ontologies for intrusion detection was made in 2003 by Undercoffer et al. \cite{undercoffer2003modeling}. They suggested an ontology that enables distributed anomaly-based host IDS sensors to contribute to a common knowledge base, which again would enable them to detect a possible attack quickly. 

Based on this, More et al. \cite{more2012knowledge} in 2012 suggested building a knowledge base with reasoning capabilities to take advantage of an extended variety of heterogeneous data sources to be able to identify threats and vulnerabilities. Their data sources suggest that data retrieved and included in the ontology is within the atomic indicators category of the CTI model.

Oltramari et al. \cite{oltramari2014building} proposed a three-layer cyber security ontology named "CRATELO" aiming at improving the situational awareness of security analysts, resulting in optimal operational decisions through semantic representation. Following the methodology of \cite{obrst2012developing}, the authors build upon existing ontologies and extend them. Specifically, CRATELO includes the top-level ontology DOLCE-SPRAY extended with the security-related middle-level ontology (SECCO), which is capable of capturing details of domain-specific scenarios, including threats, vulnerabilities, attacks, countermeasures, and assets. The low-level sub-ontology, cyber operations (OSCO), is the extension of the middle-level ontology.

Gregio et al. \cite{gregio2014ontology} suggested an ontology to address the detection of modern complex malware families whose infections involve sets of multiple exploit methods. To achieve this, they created a hierarchy of main behaviors, each one of them consisting of a set of suspicious activities. Then, they proposed an ontology that models the knowledge of malware behavior. They state that a given program behaves suspiciously if it presents one or more of the six events (main behaviors) described below, which consist of several characteristics. The events are attack launching, evasion, remote control, self-defense, stealing, and subversion. When a new set of process actions with malicious behaviors appear (input from normalized log files), the ontology can infer if an instance of suspicious execution is linked to a malware sample.

Salem and Wacek \cite{salem2015enabling} designed a data extraction tool called TAPIO (Targeted Attack Premonition using Integrated Operational data), specializing in extracting data (through the use of natural language processing) and automatically mapping that data into a fully linked semantic graph that can be accessed in real-time. Part of TAPIO is a cybersecurity ontology known as Integrated Cyber Analysis System (ICAS) that ingests extracted data (logs and events) from several sources to provide relationships across an enterprise network. The tool aims to help incident response teams connect and correlate events and actions into an ontology for automatic interpretation. ICAS is a collection of 30 sub-ontologies specializing in specific conceptual areas as part of host-based and network-based conceptual models.

Iannacone et al. \cite{iannacone2015developing} described their STUCCO ontology. Its design is based on scenarios of use by both human and automated users and incorporates data from 13 different structured data sources with different formats. The data included in the current STUCCO ontology fall into the categories of identity, TTPs, tools, and atomic indicators of the cyber threat intelligence model. Their future work mentioned extending the ontology to support STIX.

Gregio, Bonacin, de Marchi, Nabuco, and de Geus \cite{gregio2016ontology} extended the work of Gregio et al. \cite{gregio2014ontology} and introduced the malicious behavior ontology (MBO). Using SWRL rules for inferencing, MBO can detect modern complex malware families whose infections involve multiple exploit methods. In addition, these rules also apply metrics to specify whether a program is misbehaving and, specifically, how suspicious the execution of that program is. The authors state that their model can detect unknown malicious programs, even in cases where traditional security tools like antivirus cannot. This is achieved by automatically inferring suspicious executions in monitored target systems. However, the current state of the ontology has some limitations, such as performance issues, inability to detect malware in real-time, and false positives and negatives. Based on its operation, MBO can provide indicators of compromise for malware.

Fusun et al. suggested ontologies like attacks, systems, defenses, missions, and metrics for quantifying attack surfaces \cite{fusunusing}. Their Attack Surface Reasoning (ASR) allows defenders to explore trade-offs between cost and security when deciding on their cyber defense composition. ASR is mainly modeled after the Microsoft STRIDE \cite{shostack2014threat} threat modeling framework, which categorizes security threats into 6 categories.

%Security Metrics Ontology 
As part of their study on using security metrics for security modeling, Pendelton et al. suggested the Security Metric Ontology \cite{pendleton2016survey}. The ontology includes four sub-ontologies: vulnerability, attack, situations, and defense mechanisms, and describes the relationship between them. The terminology used is somewhat different than that of known taxonomies, and their aim at modeling metrics is more prominent than that of analysis and reasoning.
The ontology is published on GitHub\footnote{https://github.com/marcusp46/security-metrics-ontology}.

The Unified Cybersecurity Ontology was suggested by Syed et al. \cite{syed2016uco} in 2016. It serves as a backbone for linking cyber security and other relevant ontologies. There are mappings to aspects of STIX and references to CVE, CCE, CVSS, CAPEC, STUCCO, and KillChain. The concepts are loosely connected at a very high level, and the lack of OWL expressivity decreases the reasoning capabilities of the ontology. In addition, our analysis indicates that the use of domain and range restrictions would result in incorrect classifications (inferred axioms) while running a reasoner. The ontology is published on GitHub\footnote{https://github.com/Ebiquity/Unified-Cybersecurity-Ontology}. 

The Unified Cyber Ontology has been introduced on GitHub\footnote{https://github.com/ucoProject/uco}, without any academic publication to date and no actual RDF/OWL files yet. The model ontology is, however, interesting as it originates from the creators of STIX, which is currently the most used format for sharing threat intelligence\cite{sauerwein2017threat}. The content of that work is driven primarily by the initial base requirements of expressing cyber investigation information and is the product of input from the Cyber-investigation Analysis Standard Expression community (CASE)\footnote{https://github.com/casework/case}.

%Cyber Intelligence Ontology
Without any publication, we find the Cyber Intelligence Ontology (CIO), published only on GitHub\footnote{https://github.com/daedafusion/cyber-ontology} to be of relevance. This GitHub repository includes most of the mentioned taxonomies and sharing standards in this article, encoded in OWL. The limitation of those ontologies is that they are not connected or unified. For the aforementioned reason, we do not include CIO in the analysis and the evaluation table.

 \section{Discussion}
    
Intelligence-driven defense augments organizations' preventive, detecting, and responding capabilities by introducing a threat-informed approach to cyber security operations. The maturity of a threat intelligence program, including the analytical skills and the availability of information, determines a security team's capability to produce accurate and actionable threat intelligence \cite{rid2015attributing}\cite{johnson2016guide}.

We need unambiguous representations with sufficient expressivity and explainable relationships across concepts to leverage the benefits of ontologies and description logics in cyber threat intelligence. The analysis of the existing works confirmed that there is still a tiny focus and much work to be done to establish a comprehensive and unambiguous cyber threat intelligence ontology. A reference architecture like the one provided by our cyber threat intelligence model can be used as a conceptual blueprint to support the development of a comprehensive cyber threat intelligence ontology that is modular, extensible, and adaptive. 

Ontologies should be modular and extensible, allowing replacing or integrating with other domain-focused ontologies to build more holistic ones (per the organizations' use cases) for an augmented representation regarding a domain of interest. In the ontologies evaluated, we identified that the lack of OWL expressions is common. Expressions make ontologies powerful by encoding domain expertise and making implicit knowledge explicit through reasoning. Using the encoded knowledge, a reasoner can infer new information from the existing asserted information at machine speed, introducing a form of automation.

Furthermore, we cannot ignore mentioning the limited taxonomy encodings and integrations and the missing interconnections between those taxonomies and existing ontologies for establishing more standardized (interoperable) and expressive unambiguous representations. Examples include taxonomies for threat actor motivations, goals, and types. Standardizing and utilizing taxonomies is essential when rich and seamless querability is desired.

Overall, a cyber threat intelligence ontology should be able to formulate a knowledge graph with rich historical, present, and inferred information expressed in a meaningful and explicable way so that analysts can perform their analytical tasks assisted by automated reasoning, perform assessments, and address their knowledge gaps.

\begin{table*}[htb!]
\caption{Evaluation of taxonomies, sharing standards, and ontologies}
\centering
\label{tab:table}
\begin{adjustbox}{width=\textwidth, angle=90, scale=1.30}
\renewcommand{\arraystretch}{2.0}%
\begin{tabular}{c|c|c|c|c|c|c|c|c|c|c|c}
    
   \textbf{}   & \textbf{} & \textbf{Identity} & \textbf{Motivation} &  \textbf{Goal} & 	\textbf{Strategy} & \textbf{TTP} &\textbf{Tool} & \textbf{IOC} & 	\textbf{Atomic Indicator} & \textbf{Target} & \textbf{COA}\\ 
   
    \hline
   \textbf{Taxonomies} & TAL \cite{casey2007threat} & *  &&&&&&&&\\
   \hline
    \textbf{} & Threat Agent Motivation \cite{casey2015understanding} & *  &*&&&&&&&\\
   \hline
    \textbf{} & CVE \cite{CVE} &&&&&&&&*& \\
    \hline
    \textbf{} & NVD \cite{NVD} &&&&&&&&*& \\
    \hline
    \textbf{} & CPE \cite{CPE} &&&&&&&&*& \\
    \hline
    \textbf{} & CWE \cite{CWE} &&&&&*&&&*&&*\\
    \hline
    \textbf{} & CAPEC \cite{CAPEC} &&&&&*&&*&&&*\\
    \hline
    \textbf{} & ATT\&CK \cite{ATTCK}&*&&&&*&*&&&\\
    \hline
    \textbf{} & CVSS \cite{CVSS} &&&&&&&&*& \\
    \hline
    \textbf{} & CWSS \cite{CWSS} &&&&&&&&*&\\

% \textbf{} & &&&&&&&&&\\
    
    \hline
%%%%%%%%%%%%%%%%%%%%%%%%%%   
\hline
\textbf{Sharing Standards} & STIX 1 \cite{Barnum2012-STIX}  &*&*&\makecell{*\\(Intended Effect:taxonomy)}&\makecell{*}&*&*&*&*&*&*\\
\hline
 \textbf{} & STIX 2 \cite{STIX2} &*&*&\makecell{*\\(Objectives:string) }&\makecell{*}&*&*&*&*&*&*\\
\hline
 \textbf{} & MAEC \cite{MAEC} &&&&&&&*&&\\
    \hline    
      \textbf{} & OpenIOC \cite{OpenIOC} &&&&&*&*&*&*&\\
    \hline
%%%%%%%%%%%%%%%%%%%%%%%%%%
 %\textbf{} & &&&&&&&&&\\
    \hline
    
 \textbf{Ontologies} & Fenz \& Ekelhat (2009) \cite{fenz2009formalizing} &&&&&&&&*&\\
\hline

 \textbf{} & Wang \& Guo (2009) - OVM \cite{wang2009ovm} &&&&&*&&&*&&*\\
    \hline  
    
\textbf{} & Orbst et al. (2012) \cite{obrst2012developing}  &*&&&&&*&&*&*\\
    \hline
    
\textbf{} & More et al. (2012) \cite{more2012knowledge}  &&&&&*&&&*&\\
    \hline
 
\textbf{} & Oltramari et al. (2014) - CRATELO \cite{oltramari2014building}  &*&&&&*&&&*&*\\
    \hline
    
\textbf{} & Gregio et al. (2014) \cite{gregio2014ontology} &&&&&&\makecell{*\\(malware)}&&*&\\
    \hline

 \textbf{} & Salem \& Wacek (2015) - ICAS \cite{salem2015enabling}  &&&&&*&&&*&\\
    \hline
 
  \textbf{} & Iannacone et al. (2015) - STUCCO \cite{iannacone2015developing} &*&&&&*&*&&*&\\
    \hline

 \textbf{} & Gregio et al. (2016) - MBO \cite{gregio2016ontology} &&&&&&\makecell{*\\(malware)}&*&\makecell{*\\(it may provide)}&\\
    \hline
 
   \textbf{} & Fusun et al. (2015) - ASR \cite{fusunusing} &&&&&*&&&*&*\\
    \hline

\textbf{} & Pendelton et al. (2016) - Security Metrics Ontology \cite{pendleton2016survey} &&&&&*&&&&&*\\
    \hline

   \textbf{} & Syed et al. (2016) - UCO \cite{syed2016uco} &*&*&*&*&*&*&*&*&&*\\
    \hline

   \textbf{} & Unified Cyber Ontology (2016) - UCO \cite{UCOBarnum} &*&*&*&*&*&*&*&*&*&*\\
    \hline

  \end{tabular}
  \end{adjustbox}
  \bigskip
  \vspace{-18pt}
\end{table*}

\section{Conclusion}
Our study concluded that there is much work to be done before achieving a contextual and unambiguous cyber threat intelligence ontology. Barriers to overcome include little focus on dedicated ontological cyber threat intelligence efforts that can account for the strategic, operational, and tactical levels; ambiguity in defined concepts that prevents ontology integration and adoption; extensive use of prose and limited utilization of existing taxonomies that undermine the querability of the knowledge graphs and the ability to perform reasoning; lack of relationships between concepts that can support interpretation and explainability; and minimal use of ontology axioms and expressions that can be used for consistency checking and inference.

\IEEEpeerreviewmaketitle

\clearpage
	\bibliographystyle{IEEEtran}
	\bibliography{bibliography}

\end{document}